\documentclass{aa}  
\usepackage{natbib}        
\usepackage{hyperref}    
\usepackage{subfig}
\usepackage{graphicx}
\usepackage{txfonts}
\usepackage{amsmath} 
\usepackage{longtable}
\usepackage{lscape}
\begin{document}

   \title{T-type brown dwarfs in Euclid DR1: from spectral classification to physical parameters}

\author{
S.~L.~Zheng\inst{1}
\and
H.~R.~Gu\inst{1}
}

\institute{
School of Astronomy and Space Sciences, University of Chinese Academy of Sciences, Beijing 100049, P. R. China
}


  \abstract
   {The wide-area survey capability and near-infrared spectroscopic observations of the \textit{Euclid} mission provide a valuable opportunity to identify and characterize ultracool dwarfs. In particular, the NISP slitless spectra cover prominent molecular absorption features of T-type brown dwarfs (T-BDs), enabling their identification and atmospheric characterization over a large survey area.}
   {We aim to identify T-BDs in the \textit{Euclid} Data Release 1 (DR1), determine their spectral types and atmospheric parameters, and investigate their fundamental physical properties, including mass, age, radius, and distance.}
   {We select candidates using morphological criteria, near-infrared colour cuts, spectral fitting, and visual inspection of the multi-band images. Spectral types are determined through comparison with standard spectra using the SPLAT toolkit. Atmospheric parameters are independently determined using two model grids, Sonora Elf Owl and LOWZ, through Bayesian inference with MCMC sampling. The resulting atmospheric parameters are subsequently combined with the evolutionary models to estimate masses, ages, and radii.}
   {We identify 141 T-BDs in \textit{Euclid} DR1, including 134 newly identified objects. A comparison of the two atmospheric model grids reveals that effective temperature is the most consistent parameter, with a mean absolute difference of 94 K, whereas larger discrepancies are found for surface gravity, metallicity, and vertical mixing. The derived mass--radius distribution exhibits a clear decrease in radius with increasing mass, while the inferred ages show an overall tendency for older objects to have smaller radii. We also identify 21 relatively blue T-BDs with $I_{\rm E}-H_{\rm E}<2.5$. Their SEDs and colour distribution, including the overlap with previously reported white dwarf-brown dwarf (WD+BD) candidates, suggest that these sources may be unresolved WD+T-BD systems. We further compare distances obtained from empirical spectral type--absolute magnitude relations and MCMC fitting, finding a mean absolute difference of approximately 37 pc. In addition, three sources exhibit nearby objects in the \textit{Euclid} multi-band images and are identified as potential binary or multiple-system candidates.}
   {Our results demonstrate the potential of \textit{Euclid} DR1 for expanding the known population of T-BDs and investigating their atmospheric and evolutionary properties. The comparison between atmospheric models highlights the different levels of reliability among the derived parameters, with effective temperature providing the most robust constraint. The identified potential binary candidates also provide promising targets for future astrometric and spectroscopic follow-up observations.}


   \maketitle

\section{Introduction}

Brown dwarfs (BDs) bridge the mass gap between stars and planets, with masses typically ranging from $0.01$ to $0.08\,M_{\odot}$ \citep{burrows2001theory,meisner2021new}. Because their core temperatures are insufficient to sustain stable hydrogen fusion, they continuously cool and fade over time, making old and low-mass brown dwarfs increasingly difficult to detect \citep{chabrier2005review,davy2012further}. Their spectra exhibit prominent molecular absorption features, including water vapour ($\mathrm{H_2O}$), methane ($\mathrm{CH_4}$), and ammonia ($\mathrm{NH_3}$). These characteristics make brown dwarfs valuable laboratories for investigating the physical and chemical processes in cool atmospheres, including atmospheric chemistry and cloud formation \citep{cushing2011discovery,tu2024physical}.

Brown dwarfs are generally divided into three major spectral classes according to their infrared spectral characteristics. From higher to lower temperatures, these classes are L-type brown dwarfs, with effective temperatures of approximately 
$1200\,\mathrm{K} \lesssim T_{\mathrm{eff}} \lesssim 2300\,\mathrm{K}$; T-type brown dwarfs (T-BDs), with $500\,\mathrm{K} \lesssim T_{\mathrm{eff}} \lesssim 1200\,\mathrm{K}$; and Y-type brown dwarfs, with $T_{\mathrm{eff}} \lesssim 500\,\mathrm{K}$ \citep{cushing2011discovery,burgasser2006unified,burrows2003beyond,kirkpatrick2021field}. These cool objects emit most of their radiation at near- and mid-infrared wavelengths. Infrared telescopes are therefore powerful tools for detecting and characterizing brown dwarfs. However, due to their intrinsically low luminosities, most known brown dwarfs are confined to the solar neighbourhood \citep{tu2025large}.


Over the past few decades, large-scale surveys such as the Sloan Digital Sky Survey \citep[SDSS;][]{york2000sloan}, the Two Micron All Sky Survey \citep[2MASS;][]{skrutskie2006two}, the Wide-field Infrared Survey Explorer \citep[WISE;][]{wright2010wide}, and the Spitzer Space Telescope \citep{werner2004spitzer} have greatly expanded the known brown-dwarf population, from L dwarfs to the coolest T and Y dwarfs. More recently, the James Webb Space Telescope (JWST), with its high sensitivity and broad near- to mid-infrared wavelength coverage, has made it possible to detect and characterize much fainter and colder brown dwarfs at considerably larger distances \citep{beiler2024precise,suarez2025diversity}. For example, \citet{tu2025large} identified two brown dwarfs at distances greater than $5\,\mathrm{kpc}$ using archival JWST/NIRSpec data, making them among the most distant brown dwarfs currently known in the Milky Way. In addition, the recently launched SPHEREx mission provides continuous spectrophotometric coverage from $0.75$ to $5\,\mu\mathrm{m}$ over the entire sky. Its rapid and homogeneous survey capability has enabled the construction of a large spectrophotometric sample of ultracool dwarfs (UCDs), providing an important basis for population-level studies of their atmospheric and fundamental properties \citep{tu2026spherex,rustamkulov2026spherex,gagne2026spherex}.

In addition, the \textit{Euclid} space telescope provides a unique opportunity to identify and characterize brown dwarfs through its combination of wide-area optical and near-infrared imaging and slitless near-infrared spectroscopy \citep{jahnke2025euclid}. The combination of its wide sky coverage and deep sensitivity enables the detection of faint brown dwarfs over a much larger volume than previous surveys. In a single observation, \textit{Euclid} simultaneously provides one optical ($I_\mathrm{E}$) band, three near-infrared ($Y_\mathrm{E}$, $J_\mathrm{E}$, and $H_\mathrm{E}$) photometric measurements, and low-resolution ($R\sim500$) NISP slitless spectra covering $1.2$--$1.9\,\mu$m \citep{laureijs2010euclid}. This continuous near-infrared spectral coverage is particularly well suited for detecting the broad molecular absorption features characteristic of T-BDs \citep{kiwy2026search}.

Following the public release of Euclid Q1 in March 2025, a series of studies has demonstrated the capability of Euclid identify and characterise ultracool dwarfs. \citet{2024A&A...686A.171Z} first showed that the spectral resolution of Euclid/NISP is sufficient for the reliable
classification of UCDs. Spectroscopic and photometric searches were subsequently performed using Q1 data \citep{2026A&A...711A..37E,2025ApJ...991...84D}. In addition, \citet{2026A&A...711A..37E} spectroscopically confirmed 33 new UCDs, while \citet{2025AJ....170..360Z} identified 25 high-quality M6--L9 candidates using machine-learning techniques. Most recently,
\citet{kiwy2026search} combined Euclid colours, spectra, and imaging data to identify 15 mid-to-late T-dwarf candidates. Overall, previous studies based on the \textit{Euclid} Q1 data have confirmed a total of 18 T-BDs.

The first major Euclid Data Release (DR1) \citep{euclidcollaboration2026eucliddatarelease1}, whose first stage (DR1-Foundation) was released in November 2026, provides imaging and slitless spectroscopic observations over approximately $1900\,\mathrm{deg}^{2}$ \citep{euclidcollaboration2026eucliddatarelease1}, greatly expanding the volume accessible for brown dwarf studies. The combination of its wide sky coverage, deep near-infrared photometry, and slitless spectroscopy is expected to reveal a
substantially larger population of ultracool dwarfs, particularly late-type brown dwarfs. Recent studies have demonstrated the potential of \textit{Euclid} DR1 for brown-dwarf studies, including searches for UCDs in open clusters and red L-dwarf candidates in the wide and deep fields \citep{martín2026eucliddr1legacyscience,lodieu2026eucliddr1legacyscience}. However, a systematic investigation of T-BDs using Euclid DR1 data, particularly through a large-scale search for these objects, remains unexplored.

In this work, we utilise the Euclid DR1 data to search for T-BDs. Candidate sources are first selected as point sources using morphological criteria and near-infrared
colour--colour diagrams. Their Euclid NISP slitless spectra are then compared with spectral templates to identify T-BDs and determine their spectral types. Then, atmospheric and evolutionary models are employed to derive their physical
properties, including effective temperature, surface gravity, mass, radius, age, and distance.
In addition, we identify a population of T-BDs with unusual colours and investigate their nature through multi-band SEDs.

The paper is organized as follows. \hyperref[sec:sample]{Sect.~2} describes the \textit{Euclid} DR1 data and the construction of the T-BD candidate sample. \hyperref[sec:methods]{Sect.~3} presents the spectral classification, atmospheric-model fitting, and derivation of fundamental physical parameters. \hyperref[sec:Results]{Sect.~4} presents the main results, including the population with unusual blue colours, the identified T-BDs and their atmospheric and evolutionary properties. \hyperref[sec:Discussion]{Sect.~5} discusses the comparison with previous \textit{Euclid} studies, the consistency and limitations of the inferred parameters, and the potential binary candidates. Finally, \hyperref[sec:Summary]{Sect.~6} summarizes the main findings and discusses the prospects for future spectroscopic and astrometric follow-up observations.

\section{Sample}
\label{sec:sample}

As a first step, we isolated point-like sources following the selection criteria outlined in Table~2 of \citet{2026A&A...711A..37E}. To construct the primary sample, we applied four morphological filters: \texttt{ELLIPTICITY} $< 0.2$, \texttt{MUMAX\_MINUS\_MAG} between $-3.25$ and $-2.65~\mathrm{mag\,arcsec^{-2}}$, \texttt{FWHM} $< 1\farcs5$, and \texttt{DET\_QUALITY\_FLAG} $< 3$. A detailed description of these morphological parameters is provided in Sect.~3 of \citet{2026A&A...711A..37E}. After applying all of the above selection criteria, a total of \textbf{41,970,586} sources were retained for subsequent analysis.

\begin{figure*}
    \centering
    \includegraphics[width=\hsize]{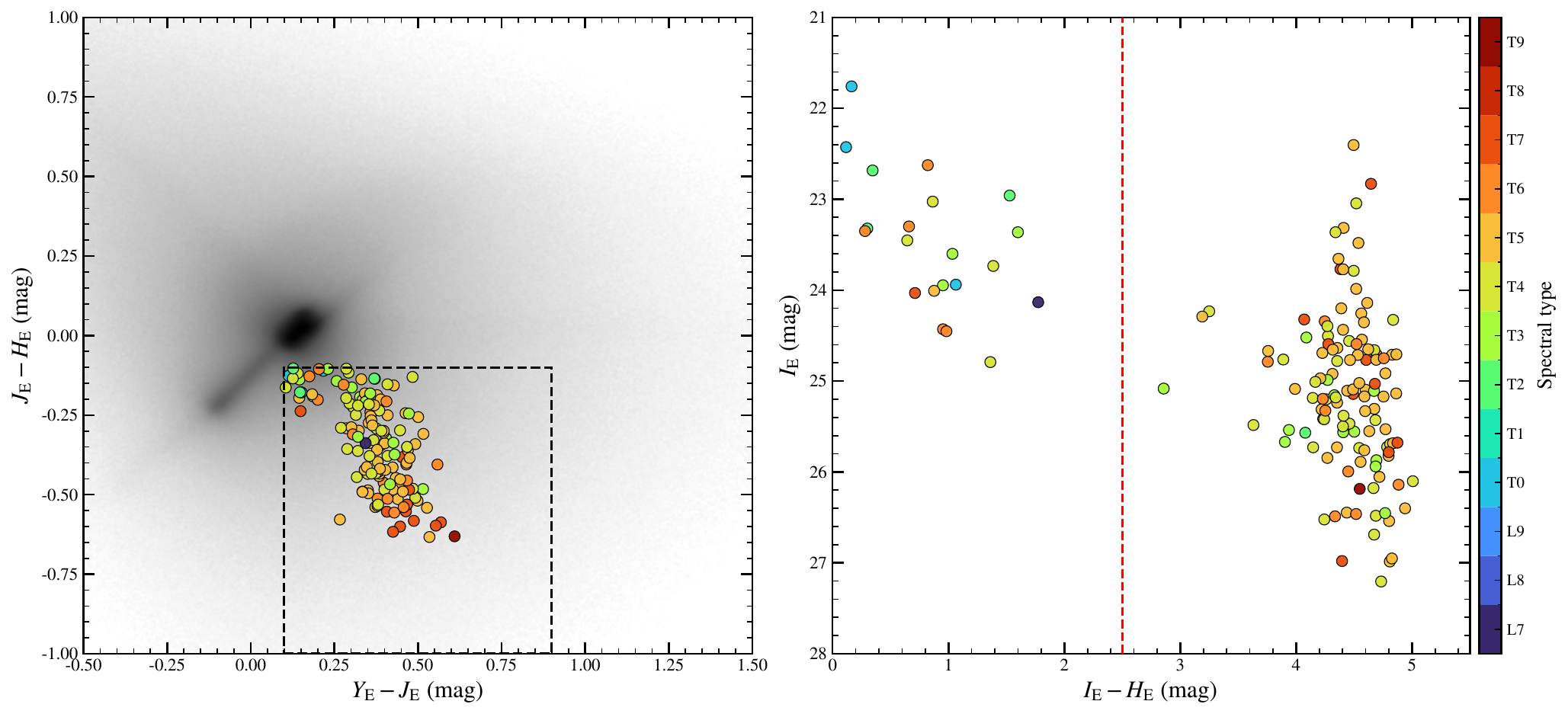}
    \caption{Euclid $Y_{\rm E}-J_{\rm E}$ versus $J_{\rm E}-H_{\rm E}$ colour--colour diagram of the selected sources. The dashed rectangle indicates the adopted photometric selection criteria. The coloured points represent our identified T-BDs, with their colours indicating the assigned spectral types, as shown by the colour bar.}
    \label{fig:color_color}
\end{figure*}

The Euclid telescope provides four different photometric measurements for each source. In this work, we adopted aperture photometry measured within an aperture radius of twice the full width at half maximum (FWHM), with magnitudes reported in the AB photometric system \citep{2026A&A...711A..37E}. To ensure the reliability of the photometric measurements, we calculated the signal-to-noise ratio (S/N) in the three NISP bands ($Y_{\rm E}$, $J_{\rm E}$, and $H_{\rm E}$) and required $\mathrm{S/N}>5$ in all bands as a photometric quality criterion. 
To select T-BDs, we further applied the colour selection proposed by \citet{kiwy2026search}, namely $-0.1 < Y_{\rm E}-J_{\rm E} < 0.9$ and $J_{\rm E}-H_{\rm E} < -0.1$. As demonstrated by \citet{kiwy2026search}, mid- to late T-BDs occupy a well-defined region in this colour space and are clearly separated from earlier-type ultracool dwarfs, making these criteria particularly effective for selecting T-BDs. After applying the above photometric quality and colour-selection criteria, a total of \textbf{1,006,565} sources remained in our sample. The left panel of Fig.~\ref{fig:color_color} shows the resulting $Y_{\rm E}-J_{\rm E}$ versus $J_{\rm E}-H_{\rm E}$ colour-colour diagram. The candidates subsequently identified as T-BDs are highlighted, with their colour indicating spectral type, thereby illustrating both their distribution within the selected colour space and the variation of spectral type across it.

It is essential that brown dwarfs be identified from their near-infrared spectra. The wavelength coverage of the Euclid NISP slitless spectra ($1.2$--$1.9\,\mu$m) encompasses the dominant molecular absorption bands of T-BDs, including the CH$_4$ absorption features at $\sim1.3$--$1.4\,\mu$m and $\sim1.6$--$1.7\,\mu$m, as well as the H$_2$O absorption bands near $\sim1.4\,\mu$m and $\sim1.9\,\mu$m. These strong molecular absorptions suppress the continuum between the $J$- and $H$-band flux peaks, producing a characteristic spectral morphology that becomes increasingly pronounced toward later T spectral types. As shown in Fig.~2 of \citet{2025ApJ...991...84D}, these features enable Euclid NISP spectra to effectively distinguish T-BDs from earlier-type ultracool dwarfs and provide reliable spectral classification. We therefore performed spectral template fitting for all candidates selected in the previous step.

Following the photometric selection, we successfully retrieved 683,168 Euclid NISP spectra and trimmed them  to the wavelength range of $1.22$--$1.88\,\mu$m to exclude low-S/N edges. For some low-S/N spectra, the spectra were smoothed using a Gaussian kernel with a width of $\sigma=1$ pixels prior to the spectral fitting. The Gaussian smoothing effectively suppresses high-frequency noise while preserving the overall continuum shape, thereby improving the robustness of the subsequent template fitting. The spectra were then fitted with the LOWZ spectral model grid (see \hyperref[sec:methods]{Sect.~3}) 

To identify the best-fitting atmospheric models, we first compared each observed Euclid spectrum with the full grid of LOWZ atmospheric templates. For each template, we calculated the reduced chi-square statistic:

\begin{equation}
\chi_r^2=
\frac{1}{N-p}
\sum_i
\frac{\left(F_i-\frac{R^2}{D^2}M_i\right)^2}
{\sigma_i^2},
\label{eq:chisq}
\end{equation}

where \(F_i\) and \(\sigma_i\) are the observed flux and its uncertainty at wavelength point \(i\), respectively, \(M_i\) is the template flux, \(N\) is the number of valid spectral data points, and \(p\) is the number of free model parameters. The scaling factor, \(R^2/D^2\), which accounts for the ratio of the object's radius to its distance, was determined analytically by minimizing \(\chi^2\):

\begin{equation}
\frac{R^2}{D^2}=
\frac{\displaystyle\sum_i \left(F_iM_i/\sigma_i^2\right)}
{\displaystyle\sum_i \left(M_i^2/\sigma_i^2\right)}.
\label{eq:scale}
\end{equation}

For each source, the template yielding the minimum reduced \(\chi_r^2\) was identified as the best-fitting template. Rather than adopting the parameters from a single template, we estimated the initial atmospheric parameters using the weighted average of the four templates with the lowest reduced \(\chi_r^2\) values. The weights were defined as inversely proportional to their respective reduced chi-square values and normalized to sum to unity.

Although reduced \(\chi_r^2\) provides a widely used statistical measure of goodness of fit by accounting for wavelength-dependent observational uncertainties, we found that it does not always adequately reflect the similarity in spectral morphology among sources with different signal-to-noise ratios. In particular, small flux uncertainties can substantially increase the chi-square contributions even when the absolute residuals between the observed and model spectra are relatively small. Consequently, spectra with high signal-to-noise ratios may exhibit relatively large reduced \(\chi_r^2\) values despite showing good agreement in their overall spectral features.

To provide a more precise assessment of spectral morphology, we therefore calculated the Pearson correlation coefficient between each observed spectrum and its best-fitting LOWZ template:

\begin{equation}
r_{\mathrm{P}}=
\frac{\displaystyle\sum_{i=1}^{N}
(F_i-\overline{F})(M_i-\overline{M})}
{\displaystyle\sqrt{
\sum_{i=1}^{N}(F_i-\overline{F})^2
\sum_{i=1}^{N}(M_i-\overline{M})^2}},
\label{eq:pearson}
\end{equation}

where \(\overline{F}\) and \(\overline{M}\) are the mean observed and best-fitting template fluxes, respectively. The Pearson correlation coefficient ranges from \(-1\) to \(1\), with values approaching unity indicating stronger agreement in spectral morphology. We used this coefficient to rank the candidates according to the similarity between their observed spectra and the corresponding best-fitting templates.

Since our primary objective is to identify T-BDs, we selected sources with best-fitting effective temperatures of \(T_{\mathrm{eff}}\leq1400\) K and Pearson correlation coefficients of \(r_{\mathrm{P}}>0.44\), corresponding to the median value for the fitted sample. These criteria were used to identify sources with both temperatures and spectral morphologies consistent with T-BDs. 

All sources satisfying these criteria were subsequently inspected visually. We examined the overall spectral morphology and the characteristic molecular absorption features of T-BDs, particularly the prominent H\(_2\)O and CH\(_4\) absorption bands, to assess whether each spectrum was consistent with classification as a T-BD and to reject obvious artefacts, poor-quality spectra, or unreliable model matches.

\begin{figure*}
    \centering
    \includegraphics[width=\hsize]{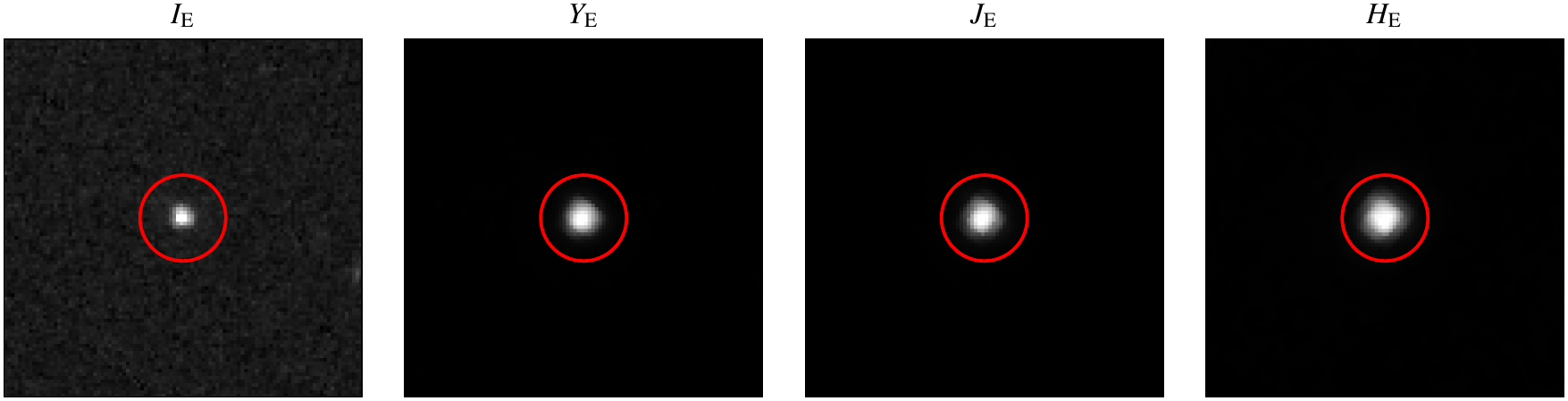}
    \caption{Euclid four-band image cutouts of Euclid id \textit{1893957385654334350}. From left to right, the panels show
the $I_{\rm E}$, $Y_{\rm E}$, $J_{\rm E}$, and $H_{\rm E}$ bands.}
    \label{fig:image_example}
\end{figure*}

In addition to the spectroscopic assessment, we examined the multi-band Euclid imaging data to further evaluate the nature of the selected sources. Using the official Euclid image cutout products\footnote{\url{https://caltech-ipac.github.io/irsa-tutorials/euclid-intro-mer-catalog/}}, we extracted 10-arcsec cutouts in the $I_{\rm E}$, $Y_{\rm E}$, $J_{\rm E}$, and $H_{\rm E}$ bands for each candidate. As illustrated in Fig.~\ref{fig:image_example}, we inspected the spatial appearance of each source to determine whether it was consistent with an unresolved point source, and checked for possible contamination from nearby objects, blending, and image artefacts. We also examined the relative source brightness across the optical and near-infrared bands, with particular attention to the expected red spectral energy distributions of cool brown dwarfs, which are generally more prominent in the near-infrared than in the $I_{\rm E}$ band. These imaging checks provided an additional means of identifying potentially contaminated or spurious candidates.

Finally, we cross-matched our candidates with previously published ultracool dwarf catalogue using their celestial coordinates \citep{best2020ultracoolsheet}. After combining the spectroscopic classification, photometric selection, and imaging inspection, we identified 141 T-BDs, of which 7 had previously been reported in the literature. The remaining 134 objects represent newly identified T-BDs. The relevant properties of all 141 T-BDs, including their coordinates, spectral types, atmospheric parameters, physical properties, and distance estimates, are summarized in Table~\ref{tab:EuclidTBD}.

\section{Methods}
\label{sec:methods}
\subsection{Spectral Classification}

We used the SpeX Prism Library Analysis Toolkit (SPLAT; \citealt{2017ASInC..14....7B}) to determine the spectral types of the T-BDs by comparing with the library of spectral standards included in SPLAT \citep{burgasser2006unified,Kirkpatrick2010,cushing2011discovery}. The classification was performed by minimizing the $\chi^{2}$ between the
observed spectrum and each standard spectrum, allowing for a multiplicative flux-scaling factor. The spectral type of the best-fitting standard was adopted as the final classification.

The uncertainties in the spectral classifications were estimated directly using the SPLAT classification procedure. For most sources, the uncertainties are approximately \(\pm1\) spectral subtype. Sources with larger classification uncertainties are indicated by one or more colons appended to their spectral types, with an increasing number of colons denoting progressively greater uncertainty. The spectral types and their corresponding uncertainties are presented in the fourth column of Table~\ref{tab:EuclidTBD}.

\subsection{Atmospheric Models}

\begin{table}[t]
\caption{Atmospheric model parameter ranges.}
\label{tab:AtmosphericModel}
\centering
\begin{tabular}{lcc}
\hline
\noalign{\smallskip}
Parameter Name & Sonora Elf Owl & LOWZ \\
\noalign{\smallskip}
\hline
\noalign{\smallskip}
$T_{\mathrm{eff}}\,[\mathrm{K}]$ & 275--2400 & 500--1600 \\
$\log g$ & 3.25--5.5 & 3.5--5.25 \\
$[\mathrm{M/H}]$ & $-1.0$--$1.0$ & $-2.5$--$1.0$ \\
$\mathrm{C/O}$ & 0.22--1.12 & 0.10--0.85 \\
$\log K_{\mathrm{zz}}$ & 2--9 & $-1$--10 \\
\noalign{\smallskip}
\hline
\end{tabular}
\end{table}

To investigate the atmospheric properties of T-BDs, we adopt two atmospheric model grids: Sonora Elf Owl \citep{wogan2025sonora} and LOWZ \citep{meisner2021new}. The parameters and their ranges are summarized in Table~\ref{tab:AtmosphericModel}. T-BDs have effective temperatures below approximately 1400~K, and their spectra are dominated by strong molecular absorption bands of H$_2$O, CH$_4$, NH$_3$, and CO. Their atmospheric structures are also strongly affected by metallicity, carbon-to-oxygen ratio (C/O), and vertical mixing. Therefore, atmospheric models capable of simultaneously accounting for these physical and chemical processes are essential for accurately characterizing T dwarf atmospheres \citep{2015ARA&A..53..279M, 2020A&A...637A..38P}.

The Sonora Elf Owl Version~2 model grid is the latest generation of the Sonora atmospheric model series, building upon the Sonora Bobcat \citep{2021ApJ...920...85M}, Sonora Cholla \citep{2021ApJ...923..269K}, and Sonora Elf Owl Version~1 \citep{mukherjee2023sonora} models. It incorporates updated molecular opacity data and adopts the chemical-equilibrium abundance for $\mathrm{PH}_{3}$, providing synthetic spectra over a broad range of atmospheric parameters. As a cloud-free atmospheric model, the Sonora Elf Owl grid is well suited for investigating cool T-BDs \citep{tu2025large}, for which silicate and iron clouds are expected to have largely settled below the photosphere following the L/T transition. Nevertheless, additional atmospheric processes may still be required to reproduce some observed spectral features \citep{2001ApJ...556..357A,2015ApJ...804L..17T}.

The LOWZ atmospheric model grid is specifically designed for low-metallicity substellar atmospheres. Compared with the Sonora Elf Owl models, LOWZ extends the metallicity coverage to much lower values, reaching $[\mathrm{M/H}]=-2.5$, making it particularly suitable for investigating metal-poor T-BDs and halo brown dwarfs \citep{2024ApJ...962..177B}. 

Although developed independently, the Sonora Elf Owl and LOWZ model grids share the same fundamental atmospheric parameters, namely $T_{\mathrm{eff}}$, $\log g$, $[\mathrm{M/H}]$, C/O, and $\log K_{\mathrm{zz}}$. The consistent parameterization enables a direct comparison between the two atmospheric models and facilitates an assessment of their systematic differences.

Before spectral fitting, both model grids were preprocessed to ensure a consistent comparison. For the Sonora Elf Owl models, only spectra with effective temperatures between 500 and 1600~K were retained to match the parameter coverage of the LOWZ grid. The high-resolution Sonora Elf Owl spectra were then convolved to the Euclid spectral resolution ($R=500$) using a Gaussian kernel. The LOWZ model library is already provided at a spectral resolution comparable to the Euclid observations, and therefore no additional spectral convolution was required.

The published LOWZ grid samples the vertical mixing parameter at only three values ($\log K_{\mathrm{zz}}=-1$, 2, and 10). To improve the sampling of the model grid, an intermediate grid point at $\log K_{\mathrm{zz}}=6$ was generated by one-dimensional interpolation in logarithmic flux space between the $\log K_{\mathrm{zz}}=2$ and $\log K_{\mathrm{zz}}=10$ models while keeping all other atmospheric parameters fixed.

\subsection{Fitting with Atmospheric Models}

The LOWZ-based initial parameter estimates obtained during the sample selection process (Sect.~\ref{sec:sample}) were adopted as the common initial values for the subsequent Bayesian analysis. Starting from this same initial parameter estimate, Bayesian inference was performed independently with the Sonora Elf Owl and LOWZ model grids to derive their respective posterior probability distributions and parameter uncertainties. Following the independent Bayesian fitting with the two atmospheric model grids, we evaluated the goodness of fit of their respective best-fitting spectra using the reduced \(\chi^2\) statistic. This statistic accounts for the wavelength-dependent observational uncertainties and provides a quantitative measure of the agreement between the model predictions and the observed spectra. The atmospheric parameters derived from the model grid yielding the lower reduced \(\chi^2\) value were adopted as the final solution for each target.

The initial parameter estimates were used to define the prior ranges for $T_{\mathrm{eff}}$, $\log g$, and the flux-scaling parameter. Bounded uniform priors were adopted for all free parameters. Specifically, $T_{\mathrm{eff}}$ and $\log g$ were allowed to vary within $\pm200$~K and $\pm1.0$~dex, respectively, around the initial solution. For the remaining atmospheric parameters, including $[\mathrm{M/H}]$, C/O, and $\log K_{\mathrm{zz}}$, uniform priors covering the full parameter ranges of the corresponding model grid were adopted. The flux-scaling parameter, $\log(R^{2}/D^{2})$, was assigned a uniform prior within $\pm0.2$~dex of the initial scaling factor.

Instead of sampling directly on the discrete atmospheric model grid, we constructed a local five-dimensional interpolator for each target using the \texttt{RegularGridInterpolator} algorithm. The local interpolation grid was built only within the parameter ranges defined by the adopted priors, thereby avoiding the computational expense of constructing an interpolator over the entire model library. Linear interpolation was simultaneously performed in the five atmospheric parameters, namely $T_{\mathrm{eff}}$, $\log g$, $[\mathrm{M/H}]$, C/O, and $\log K_{\mathrm{zz}}$. Since a small fraction of the atmospheric model grid is incomplete, some grid nodes required for constructing the local interpolator are unavailable. In these cases, the missing grid points are replaced by the nearest available template. This procedure transforms the discrete model grid into a continuous parameter space, allowing synthetic spectra to be evaluated at arbitrary parameter combinations during the sampling process. Compared with direct sampling on the original model grid, the local multidimensional interpolator significantly improves the computational efficiency while providing smooth posterior distributions and reducing quantization effects introduced by the finite model-grid spacing.

Posterior sampling was performed using the affine-invariant ensemble Markov Chain Monte Carlo (MCMC) sampler implemented in the \texttt{emcee} package \citep{foreman2013emcee}. Each fit employed 32 walkers and 10\,000 sampling steps, with the first 5\,000 steps discarded as burn-in. At each MCMC step, the interpolated synthetic spectrum was scaled by the factor $R^{2}/D^{2}$ before comparison with the observed spectrum.

Assuming Gaussian observational uncertainties, the logarithmic likelihood function is given by
\begin{equation}
\ln \mathcal{L}
=
-\frac{1}{2}
\sum_i
\left[
\frac{\left(F_{i}-\frac{R^2}{D^2}M_{i}\right)^2}
{\sigma_i^2}
+
\ln\left(2\pi\sigma_i^2\right)
\right],
\end{equation}

\begin{figure}
    \centering
    \includegraphics[width=\hsize]{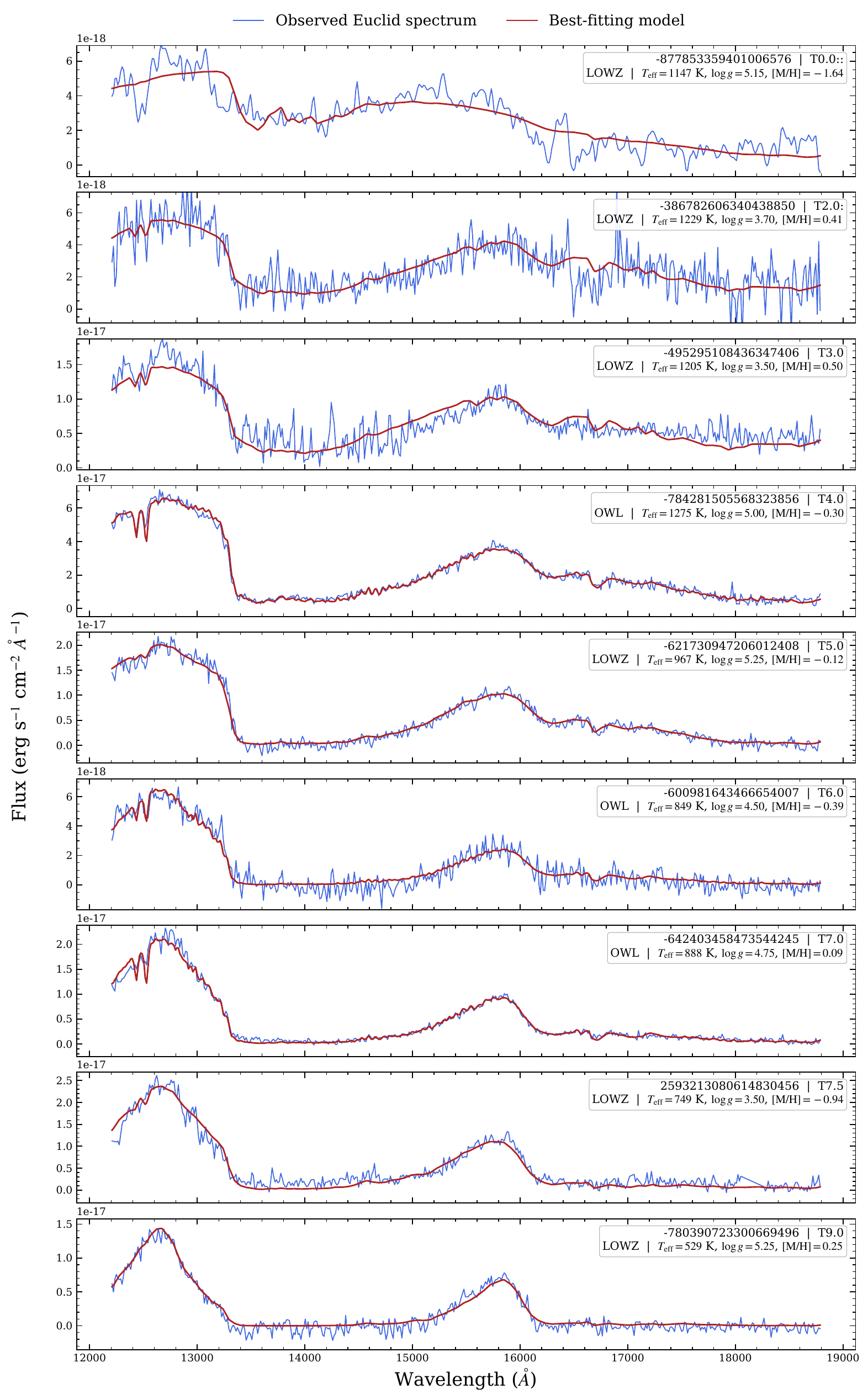}
\caption{Spectral sequence of T-BDs identified in Euclid DR1. Each panel presents an observed Euclid NISP spectrum (blue) together with its best-fitting atmospheric model (red), covering the available spectral subclasses from T0 to T9 in our sample. The panels are arranged in order of increasing spectral subtype. The object identifier, adopted spectral type, best-fitting model grid (LOWZ or Owl), and corresponding atmospheric parameters are indicated in each panel. This sequence provides a spectral atlas of the T-BDs identified in this work.}
    \label{fig:Four_TBD_Best_Model_Spectra}
\end{figure}

The posterior distributions were characterized using the 16th, 50th, and 84th percentiles, corresponding to the median values and the $1\sigma$ uncertainties of each atmospheric parameter. The parameter vector with the maximum posterior probability was adopted to generate the best-fitting synthetic spectrum. The reduced $\chi^{2}$ value calculated from this spectrum was then used to compare the performances of the Sonora Elf Owl and LOWZ model grids. The atmospheric parameters obtained from the model grid with the smaller reduced $\chi^{2}$ value were adopted as the final solution for each target. The final atmospheric parameters are shown in Table~\ref{tab:EuclidTBD}. The best-fit spectral atlas is presented in Fig.~\ref{fig:Four_TBD_Best_Model_Spectra}, ordered from top to bottom by progressively later spectral types.

\subsection{Evolutionary Model}

The atmospheric parameters derived from the Bayesian spectral fitting were subsequently combined with evolutionary models to infer the physical properties of the T-BDs, including their masses, ages, and radii. We adopted the evolutionary models of \citet{chabrier2023impact}, which utilise ATMO2020 atmospheric models \citep{2020A&A...637A..38P} and incorporate an updated equation of state for dense hydrogen–-helium mixtures, including H–He interactions, in the evolutionary calculations of very low-mass substellar objects.

The evolutionary tracks were read from all available model files and combined into a single two-dimensional grid in mass and age. Each grid point contains the corresponding effective temperature ($T_{\mathrm{eff}}$), surface gravity ($\log g$),   and radius ($R$). To efficiently evaluate the evolutionary model at arbitrary parameter values, a two-dimensional interpolator was constructed using the \texttt{LinearNDInterpolator} algorithm in the $(M,\log_{10}(\mathrm{Age/yr}))$ parameter space. When linear interpolation was not available outside the convex hull of the model grid, the nearest-neighbour interpolation provided by \texttt{NearestNDInterpolator} was adopted to ensure numerical stability.

The atmospheric parameters derived in the previous section were used as constraints on the evolutionary models. Specifically, the median \(T_{\mathrm{eff}}\) and \(\log g\) values, together with their corresponding uncertainties, were adopted to infer the mass and age of T-BDs. Assuming uniform priors in mass and logarithmic age, the posterior distributions of these evolutionary parameters were sampled with MCMC using 100 walkers evolved for 5000 steps. The burn-in length was set to twice the integrated autocorrelation time when this quantity could be reliably estimated; otherwise, the first half of the chains was discarded.

For a given parameter vector $(M,\log \mathrm{Age})$, the interpolator predicts the corresponding $T_{\mathrm{eff}}$ and $\log g$. Assuming Gaussian uncertainties, the logarithmic likelihood is

\begin{align}
\ln \mathcal{L}
=
-\frac{1}{2}
\Bigg[
&\frac{\left(T_{\mathrm{eff,obs}}-T_{\mathrm{eff,mod}}\right)^2}
{\sigma_{T_{\mathrm{eff}}}^{2}}
+
\frac{\left(\log g_{\mathrm{obs}}-\log g_{\mathrm{mod}}\right)^2}
{\sigma_{\log g}^{2}} \nonumber \\
&+ \ln\left(2\pi\sigma_{T_{\mathrm{eff}}}^{2}\right)
+
\ln\left(2\pi\sigma_{\log g}^{2}\right)
\Bigg],
\end{align}

where $T_{\mathrm{eff,obs}}$ and $\log g_{\mathrm{obs}}$ are the atmospheric parameters derived from the spectral fitting, while $T_{\mathrm{eff,mod}}$ and $\log g_{\mathrm{mod}}$ are predicted by the evolutionary model.

\begin{figure}
    \centering
    \includegraphics[width=\hsize]{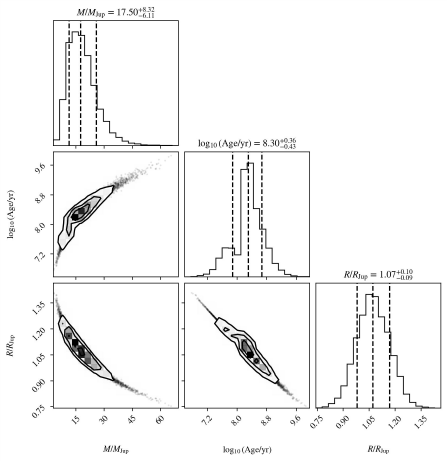}
    \caption{Posterior distributions and correlations of the derived mass, age, and radius for a representative T-BD, obtained from the evolutionary-model inference.}
\label{fig:corner_-198166084591891264}
\end{figure}

At each MCMC step, the sampled mass and age were used to interpolate the corresponding radius from the evolutionary model. The posterior distributions of these physical parameters were summarized using the 16th, 50th, and 84th percentiles, corresponding to the median values and their $1\sigma$ uncertainties. The derived mass, age, and radius are presented in Table~\ref{tab:EuclidTBD}.
An example of the corner plot showing the posterior distributions and correlations among the fitted physical parameters is presented in Fig.~\ref{fig:corner_-198166084591891264}.

\section{Results}
\label{sec:Results}

\subsection{Potential White Dwarf--T-BD Binary Candidates}

As shown in the right panel of Fig.~\ref{fig:color_color}, the colour--magnitude diagram reveals an interesting bimodal distribution of our T-BDs, with the sources separated by the red dashed line at $I_{\rm E}-H_{\rm E}=2.5$. The two groups exhibit noticeably different photometric properties: sources with $I_{\rm E}-H_{\rm E}<2.5$ are generally bluer and brighter, whereas those with $I_{\rm E}-H_{\rm E}\geq2.5$ are redder and fainter. Interestingly, a similar separation is also found when using the template-fitting photometry, suggesting that this feature is unlikely to arise solely from the adopted aperture-photometry measurements. We therefore re-examined the near-infrared spectra and multi-band images of the blue sources. Their spectral features remain consistent with the characteristic molecular absorption signatures of T-BDs, supporting their classification as T-type brown dwarfs.

Brown dwarfs generally exhibit red optical-to-infrared colours because their spectral energy distributions peak at near-infrared wavelengths. Previous photometric searches for ultracool dwarfs with \textit{Euclid} have therefore adopted red-colour selection criteria, such as $I_{\rm E}-Y_{\rm E}>2.5$ \citep{2026A&A...711A..37E}. However, the presence of a distinct population of relatively blue sources in our sample is particularly intriguing. One possible explanation is that these objects are unresolved white dwarf--T-BD binary systems. In such systems, the white dwarf could contribute a substantial fraction of the optical emission, shifting the combined optical-to-infrared colour towards the blue, while the T-BD remains the dominant contributor at near-infrared wavelengths. This composite emission could therefore account for the unusual colours observed in the blue population.

To investigate this possibility, we first cross-matched our sample with the Galaxy Evolution Explorer (GALEX; \citealt{2005ApJ...619L...1M}) catalogue to search for ultraviolet counterparts that could provide additional evidence for a hot white dwarf component. However, only one source has a counterpart within an angular separation of $2\arcsec$, and its ultraviolet detection has a very low S/N. Consequently, this association cannot be regarded as reliable evidence for ultraviolet emission from the target, and the GALEX data do not currently provide a meaningful constraint on the presence of white dwarf companions.

\begin{figure}
    \centering
    \includegraphics[width=\hsize]{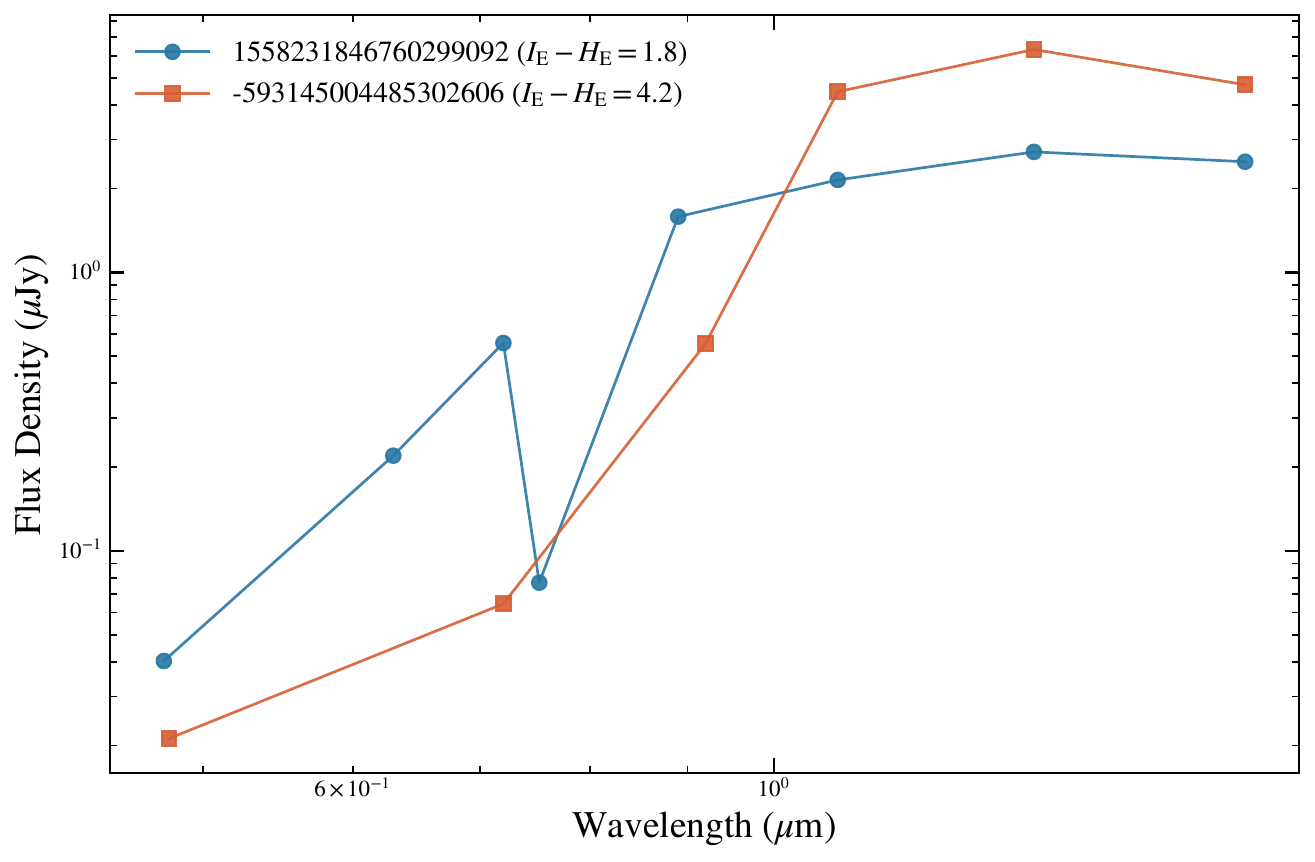}
\caption{Spectral energy distributions (SEDs) of two representative T-BDs from the two colour populations. The blue circles and line represent the source with $I_{\rm E}-H_{\rm E}=1.8$, while the orange squares and line represent the source with $I_{\rm E}-H_{\rm E}=4.2$. The blue source exhibits relatively stronger optical emission compared with its near-infrared emission, in contrast to the red source, which shows the typical near-infrared-dominated SED of a T-BD.}
\label{fig:SED_comparison_two_TBDs}
\end{figure}

We subsequently examined the available ground-based photometry incorporated into the \textit{Euclid} catalogue to investigate the spectral energy distributions (SEDs) of the blue sources. The corresponding multi-band flux densities are presented in Table~\ref{tab:blue_source_flux}. Fig.~\ref{fig:SED_comparison_two_TBDs} compares the SEDs of two representative objects: one with $I_{\rm E}-H_{\rm E}=4.2$ from the red population and another with $I_{\rm E}-H_{\rm E}=1.8$ from the blue population. The red source exhibits the expected behaviour of a typical T-BD, with relatively weak optical emission and substantially stronger near-infrared emission. In contrast, the blue source shows a pronounced enhancement in the optical bands relative to its near-infrared flux distribution. This difference suggests the presence of an additional optical emission component that is not readily explained by the expected SED of a single T-BD. The contribution of a hot unresolved companion, such as a white dwarf, therefore represents a plausible explanation for this behaviour.

Unresolved white dwarf--brown dwarf systems have been investigated in previous studies \citep{2015MNRAS.447.3218C,2020MNRAS.498...12H,2026RAA....26i5017W}. For example, \citet{2011ApJS..197...38D} presented a systematic search for infrared excesses around white dwarfs in the SDSS DR7 catalogue and identified candidate WD+BD systems through multi-wavelength SED fitting. Fig.~8 of their work presents representative SEDs of WD+BD candidates, showing the observed photometry together with the best-fitting composite WD+BD models. The resulting SEDs exhibit a characteristic distribution in which the white dwarf dominates the shorter-wavelength emission, while the brown dwarf contributes increasingly at longer wavelengths. This behaviour is qualitatively consistent with the SED morphology of the blue source shown in Fig.~\ref{fig:SED_comparison_two_TBDs}, which represents one of the most prominent examples among our 21 blue candidates. Nevertheless, the similarity in SED morphology alone is insufficient to establish the presence of a white dwarf companion, and dedicated composite-atmosphere model fitting is required to test this interpretation.

To further investigate the possibility that the blue T-BDs are unresolved WD+T-BD binary systems, we compared their colour distributions with those of previously identified WD+BD binary candidates. We selected 19 WD+BD candidates reported by \citet{2026ApJ..1001...81W}, who identified these systems through multi-wavelength photometric analysis and SED fitting. We cross-matched their Gaia DR3 source IDs with the multi-band photometric data available through the Spanish Virtual Observatory (SVO). Using the retrieved photometric measurements, we reconstructed the SEDs of these objects and performed synthetic photometry by convolving the SEDs with the \textit{Euclid} VIS, $Y$, $J$, and $H$ filter transmission curves. This allowed us to obtain their corresponding synthetic magnitudes in the \textit{Euclid} photometric system and compare them directly with our Euclid DR1 T-BD sample.

\begin{figure}
    \centering
    \includegraphics[width=\hsize]{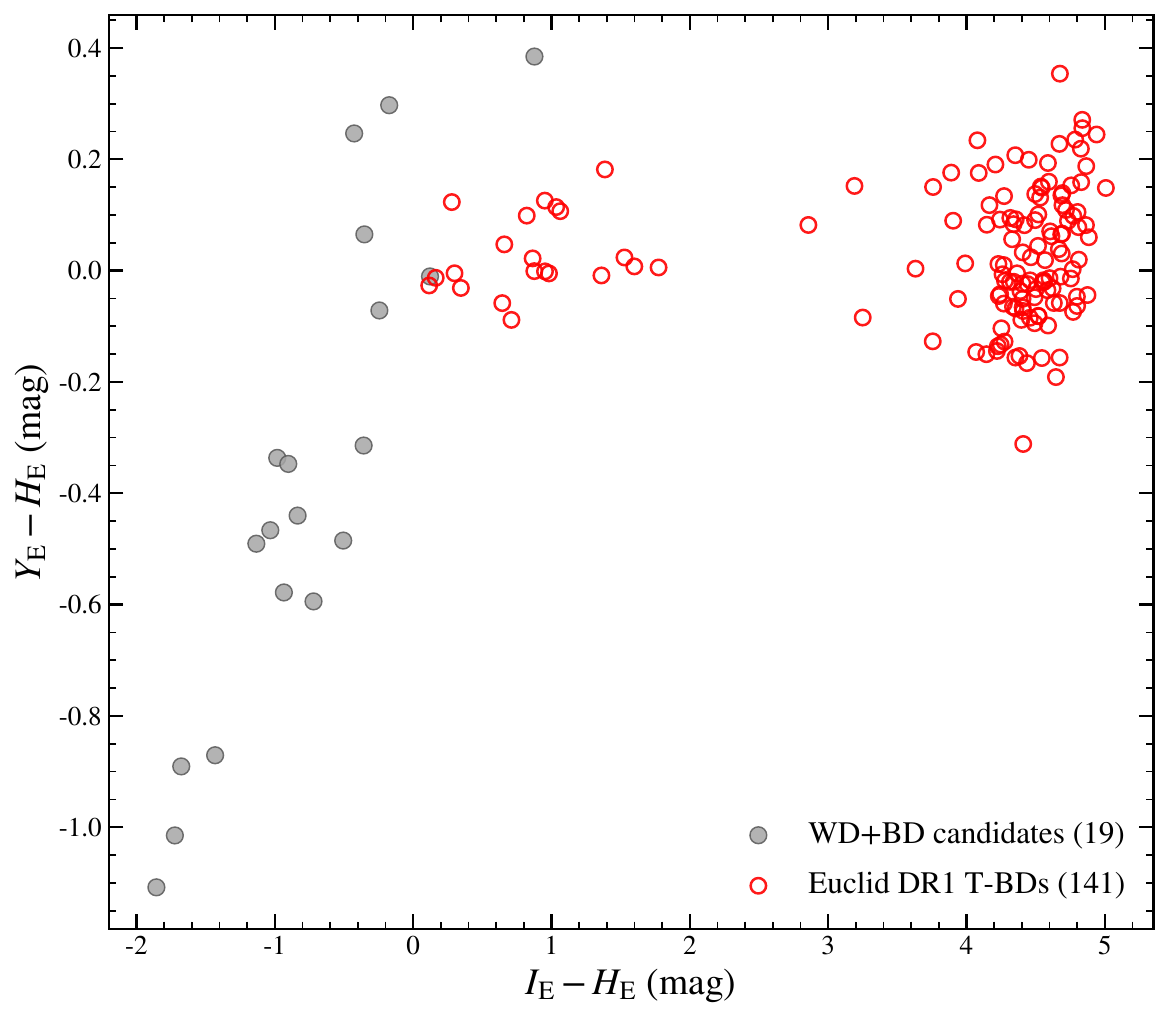}
\caption{Colour--colour diagram comparing the WD+BD candidates reported by \citet{2026ApJ..1001...81W} with the T-BDs identified in our Euclid DR1 sample. The grey filled circles represent the 19 WD+BD candidates, for which synthetic Euclid magnitudes were derived from their multi-band SEDs using the Euclid filter transmission curves. The red open circles represent the 141 T-BDs identified in our work. The horizontal and vertical axes show $I_{\rm E}-H_{\rm E}$ and $Y_{\rm E}-H_{\rm E}$, respectively.}  
\label{fig:WD_BD_color_color}
\end{figure}

Figure~\ref{fig:WD_BD_color_color} shows the resulting $(I_{\rm E}-H_{\rm E})$ versus $(Y_{\rm E}-H_{\rm E})$ colour--colour diagram. The grey filled circles represent the 19 WD+BD candidates from \citet{2026ApJ..1001...81W}, while the red open circles denote the 141 T-BDs identified in our Euclid DR1 sample. Interestingly, a subset of the WD+BD candidates occupies the bluer region of the diagram, overlapping with the distribution of our blue T-BD population. This similarity suggests that the photometric properties of at least some of our blue sources are consistent with those expected for WD+BD systems. In particular, the additional optical emission from a hot white dwarf can shift the combined optical-to-infrared colours towards the blue, providing a possible explanation for the unusual colour distribution observed in our sample.

Despite the interesting photometric evidence, the nature of these sources remains uncertain. Owing to the faintness of the targets and the depth of the \textit{Euclid} observations, multi-wavelength follow-up data are currently limited, particularly in the ultraviolet, where the contribution from a hot white dwarf would be most readily identified. The lack of reliable ultraviolet detections therefore prevents us from independently confirming the proposed companion scenario. Nevertheless, the combination of their SEDs, T-type spectral characteristics, and image data makes these objects promising candidates for unresolved WD+T-BD systems. Future ultraviolet observations will be crucial for reconstructing the complete SEDs of these sources over a broader wavelength range, thereby providing stronger constraints on the possible white dwarf contribution and helping to determine whether these objects are unresolved WD+T-BD systems.

\subsection{Comparison of Atmospheric Parameters of Two Models}
\label{sec:comparison_parameters}

\begin{figure*}[htbp]
\centering
\includegraphics[width=0.8\textwidth]{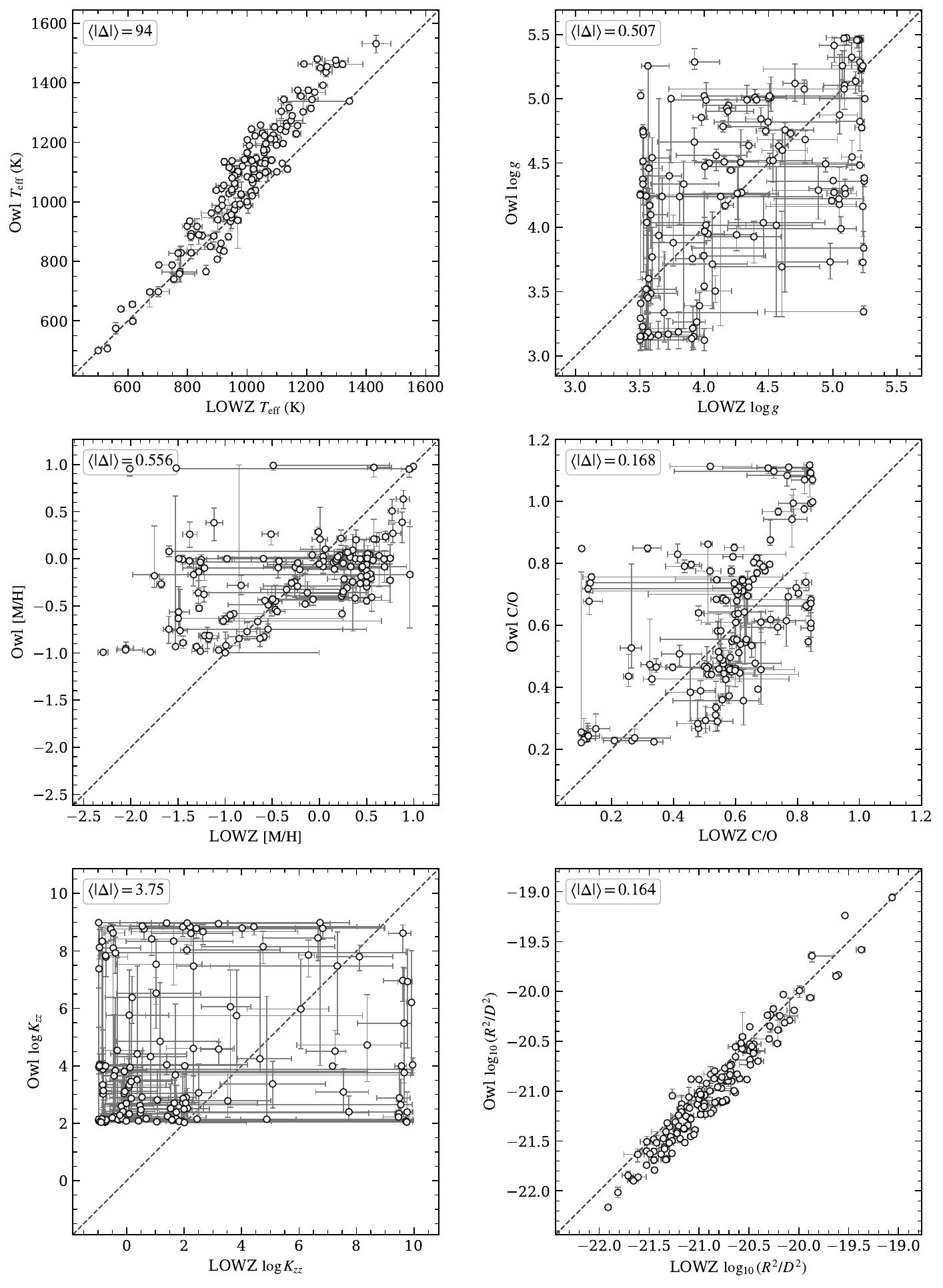}
\caption{Comparison of the atmospheric parameters derived from the Owl and LOWZ model grids for the T-BDs. From top left to bottom right, the panels compare $T_{\rm eff}$, $\log g$, [M/H], C/O, $\log K_{zz}$, and $\log_{10}(R^2/D^2)$. The dashed line in each panel indicates the one-to-one relation, and the mean absolute difference, $\langle|\Delta|\rangle$, between the two model estimates is shown in the upper-left corner. Error bars represent the corresponding parameter uncertainties.}
\label{fig:comparison_parameter}
\end{figure*}

Fig.~\ref{fig:comparison_parameter} presents a comparison of the atmospheric parameters derived from the Owl and LOWZ models for our T-BDs. The dashed lines indicate the one-to-one relations, while the mean absolute differences between the two sets of estimates are given in each panel. Overall, the comparison reveals different levels of consistency among the atmospheric parameters, suggesting that their constraints are sensitive to both the spectral information available and the underlying model assumptions.

The effective temperature ($T_{\rm eff}$) shows the best agreement between the two model grids, with a mean absolute difference of only $94\,\mathrm{K}$. Most sources lie close to the one-to-one relation, indicating that Sonora Elf Owl and LOWZ provide broadly consistent temperature estimates. Similar agreement in $T_{\rm eff}$ between different atmospheric model grids has also been reported in previous studies of cool brown dwarfs \citep{tu2025large}. A slight systematic offset becomes apparent toward the higher-temperature end of our sample, where the Owl models tend to yield somewhat higher $T_{\rm eff}$ values than LOWZ. This offset may be related to increasing model sensitivity near the L/T transition, where condensate clouds become increasingly important and cloud-free atmospheric models can have difficulty reproducing the observed spectral morphology. Nevertheless, the overall differences remain small, suggesting that $T_{\rm eff}$ is the most robustly constrained atmospheric parameter in our analysis.

In contrast, the surface gravity ($\log g$) exhibits considerably larger discrepancies, with a mean absolute difference of 0.507~dex. The sources are widely scattered across the parameter space, with a relatively weak correspondence between the two sets of estimates. This indicates that the determination of surface gravity is more sensitive to the adopted atmospheric models. In brown dwarf spectra, gravity-sensitive information is generally encoded in pressure-dependent spectral features and the overall atmospheric structure. However, these signatures can be difficult to distinguish from the effects of effective temperature, metallicity, cloud properties, and other atmospheric parameters. In particular, \citet{2023ApJ...954...22L} compared multiple atmospheric model grids and found that, while effective temperature can be robustly recovered across different models, the inferred surface gravity exhibits significant model dependence. Consequently, degeneracies among these parameters, together with differences in the model prescriptions, can substantially affect the derived surface gravities. This also makes $\log g$ a challenging parameter to constrain reliably from near-infrared spectra alone.

The metallicity ([M/H]) shows an even larger mean absolute difference of approximately 0.556~dex. The comparison reveals substantial scatter and a systematic difference between the two model estimates, indicating that metallicity is not consistently constrained by the two atmospheric grids. In particular, the Owl estimates appear to be concentrated within a relatively narrow range, whereas the LOWZ estimates exhibit a broader distribution. The apparent concentration of sources towards relatively high metallicities should therefore be interpreted with caution. 
Previous studies have shown that metallicity estimates for brown dwarfs can be strongly affected by the adopted atmospheric model and wavelength coverage \citep{tu2024physical,tu2025large}. As illustrated by the spectral-model comparisons in Fig.~7 of \citet{2024ApJ...963...73M}, variations in metallicity affect several molecular absorption regions across the near- and mid-infrared. In addition to the metallicity-sensitive features within the \textit{Euclid} NISP wavelength range, further diagnostic information is present at longer near-infrared wavelengths around 2.0--2.5\,$\mu\mathrm{m}$, as well as in the CO$_2$ and CO absorption bands at approximately 4.2--5.0\,$\mu\mathrm{m}$, respectively. The latter bands in particular exhibit stronger spectral variations with metallicity. Because these wavelength regions fall outside the \textit{Euclid} NISP spectral coverage, the available spectra provide only limited leverage for breaking degeneracies between metallicity and other atmospheric parameters. This restricted wavelength coverage likely contributes to the relatively weak and model-dependent constraints on $[\mathrm{M/H}]$ found in our analysis.

For the carbon-to-oxygen ratio (C/O), the two models show a comparatively smaller mean absolute difference of 0.162. Most sources are distributed around the one-to-one relation, suggesting a reasonable degree of agreement between the two model predictions. This relatively good agreement may be partly attributed to the fact that the \textit{Euclid} NISP wavelength range includes prominent CH$_4$ absorption features that are sensitive to variations in C/O. As illustrated by the model comparisons in Fig.~7 of \citet{2024ApJ...963...73M}, the spectral separation between different C/O values is particularly evident around 1.6--1.8\,$\mu\mathrm{m}$. The presence of these C/O-sensitive CH$_4$ features therefore provides useful leverage for constraining this parameter and may explain the relatively small differences obtained between the two atmospheric model grids \citep{tu2024physical}. Nevertheless, the CH$_4$ bands are also affected by other atmospheric parameters, and the inferred C/O values may still be subject to parameter degeneracies and model-dependent systematics.

The vertical mixing parameter ($\log K_{zz}$) exhibits the largest discrepancy among the atmospheric parameters, with a mean absolute difference of approximately 3.75~dex. The estimates show substantial scatter, and many sources appear to cluster near particular regions of the parameter grids. In particular, some values lie close to the boundaries of the available model parameter space, suggesting that the fitting procedure may not always identify a well-constrained optimum within the grid. The large differences between the Owl and LOWZ estimates indicate that vertical mixing is particularly sensitive to the adopted atmospheric prescriptions and the available spectral diagnostics. Consequently, the quantitative values of $\log K_{zz}$ should be interpreted with considerable caution.

Finally, the scaling parameter, expressed as $\log_{10}(R^2/D^2)$, exhibits good agreement between the two models, with a mean absolute difference of approximately 0.164~dex. Most sources closely follow the one-to-one relation, although a slight systematic offset is visible. This agreement is expected to some extent because the scaling parameter is primarily determined by matching the overall flux normalization of the observed and model spectra. The small residual differences between the two model estimates may partly arise from the degeneracy between \(T_{\rm eff}\) and the scaling parameter, since different effective temperatures can require different values of \(R^2/D^2\) to reproduce the same observed flux level. In addition, the accuracy with which the scaling parameter can be determined depends on the quality of the observed spectrum; spectra with lower S/N generally provide weaker constraints on the overall flux normalization and therefore lead to larger uncertainties in \(\log_{10}(R^2/D^2)\).

Overall, our comparison demonstrates that the reliability of atmospheric parameter determination varies substantially from one parameter to another. The effective temperature is consistently recovered by both models and therefore represents the most reliable atmospheric parameter in our analysis. In contrast, surface gravity, metallicity, C/O, and particularly vertical mixing show larger discrepancies, highlighting the limitations imposed by spectral quality, wavelength coverage, parameter degeneracies, and differences in atmospheric model assumptions. In general, higher S/N spectra are expected to provide stronger constraints on atmospheric properties, although the intrinsic sensitivity of the available spectral features and the adopted model prescriptions remain equally important. These results emphasize that atmospheric parameters should not be interpreted as equally well constrained. For the present Euclid DR1 near-infrared spectra, we recommend placing the greatest confidence in the effective temperature estimates, while treating the remaining atmospheric parameters more cautiously.

\subsection{Evolutionary Properties: Mass--Radius Relation of Brown Dwarfs}

\begin{figure}[htbp]
\centering
\includegraphics[width=\hsize]{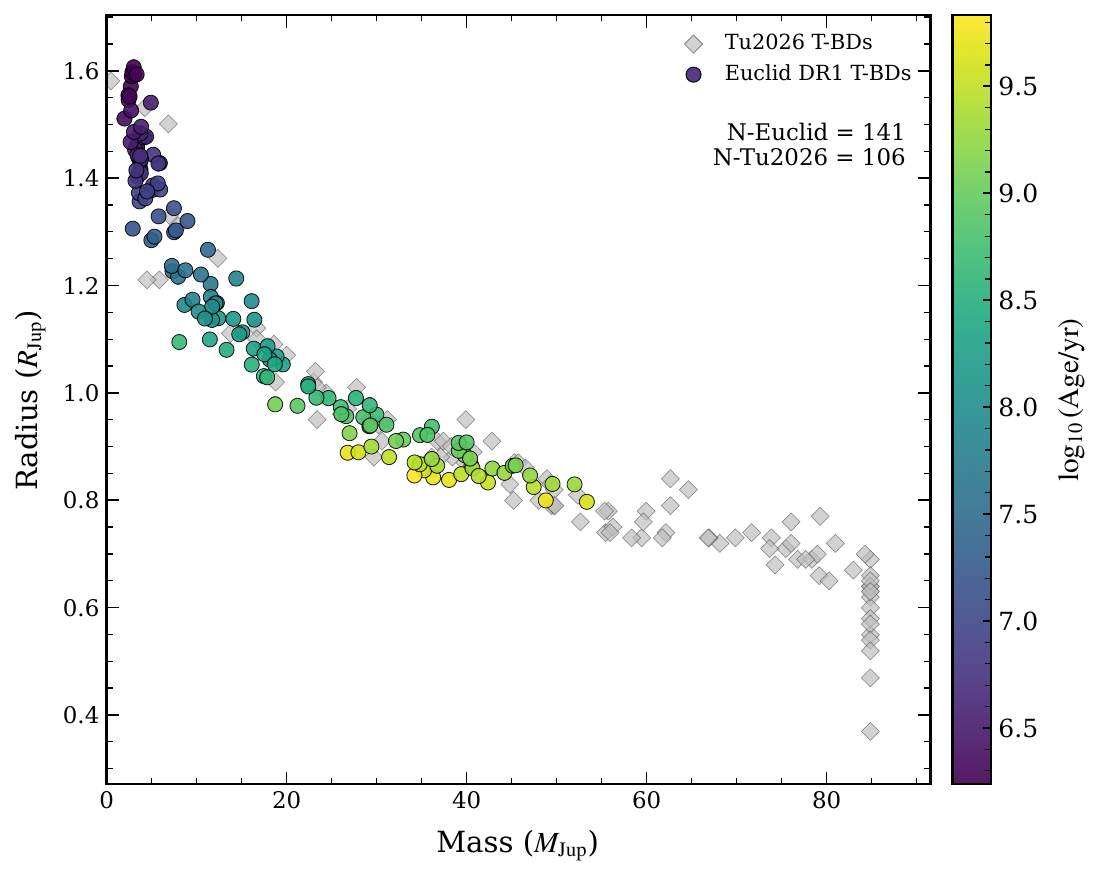}
\caption{Mass--radius distribution of the 141 T-BDs identified in this work using Euclid DR1, compared with 106 T-BDs from \citet{tu2026spherex}. The Euclid DR1 sources are shown as coloured circles, with the colour scale representing their inferred ages in $\log_{10}(\mathrm{Age/yr})$, while the Tu2026 sources are shown as grey diamonds. The combined distribution provides a broader observational context for investigating the mass--radius properties of T-BDs.}
\label{fig:T_DR1_mass_radius_age_distribution}
\end{figure}

Fig.~\ref{fig:T_DR1_mass_radius_age_distribution} presents the mass--radius distribution of our T-BDs, together with 106 T-BDs from the SPHEREx Ultracool Dwarf spectrophotometric Atlas (SUDA; \citealt{tu2026spherex}, hereafter Tu2026). The Euclid DR1 sample consists of 141 sources, shown as coloured circles, with the colour scale representing their inferred ages in $\log_{10}(\mathrm{Age/yr})$, whereas the Tu2026 sample is represented by grey diamonds. The age distribution of our Euclid DR1 sample also reveals a tendency for older objects to occupy regions with smaller radii, while younger objects generally exhibit larger radii.

A comparison between the two samples reveals differences in their distributions across the mass--radius plane. The Euclid DR1 sources are predominantly concentrated in the lower-mass region, whereas the Tu2026 sample extends toward higher masses and smaller radii. These differences may partly arise from the different approaches adopted to constrain the evolutionary properties. In our analysis, the masses, ages, and radii are inferred by fitting the atmospheric parameters, particularly $T_{\rm eff}$ and $\log g$, to evolutionary models. However, as discussed in Sect.~\ref{sec:comparison_parameters}, the spectroscopically derived $\log g$ values show considerable model dependence and are subject to degeneracies with other atmospheric parameters, potentially introducing additional uncertainties into the inferred evolutionary properties. In contrast, Tu2026 uses $T_{\rm eff}$ and radii derived from spectrophotometric flux scaling combined with literature parallaxes to constrain the evolutionary models. The different observational constraints may therefore contribute to the discrepancies between the two distributions. In addition, the mass distribution of our Euclid DR1 sample is limited by the upper boundary of the adopted evolutionary model grid, which extends to approximately $78\,M_{\rm Jup}$. We did not extrapolate beyond this boundary, and consequently, our analysis does not provide mass estimates outside the available model coverage. This limitation may further contribute to the different distributions at the high-mass end. Therefore, the observed differences should be interpreted with caution, as they may reflect not only the intrinsic properties of the samples but also the different parameter constraints, associated uncertainties, and model-grid coverage.

The overall decrease in radius with increasing mass is broadly consistent with theoretical expectations for substellar objects. Unlike ordinary stars, brown dwarfs are strongly influenced by electron degeneracy pressure in their interiors. As the mass increases, the contribution of electron degeneracy becomes increasingly important, resulting in a different mass--radius relation from that of hydrogen-burning stars. The evolutionary calculations presented by \citet{2011ApJ...736...47B} demonstrate a similar decrease in radius with increasing mass across the brown dwarf regime, following a relatively broad radius maximum at lower masses.

The age distribution further provides information about the evolutionary properties of our sample. Since brown dwarfs cannot sustain hydrogen fusion in their cores, they gradually lose their internal thermal energy through radiation. Consequently, they undergo continuous cooling, leading to a decrease in radius with increasing age, particularly during the early evolutionary stages. This behaviour is consistent with the evolutionary models discussed by \citet{2011ApJ...736...47B}, which predict a systematic decrease in brown dwarf radii with age. \citet{2013ApJ...767...77S} also discussed the combined effects of mass and age on the radius distribution of brown dwarfs, highlighting the importance of electron degeneracy in determining their structural properties. Furthermore, the evolutionary calculations of \citet{2003A&A...402..701B} provide a theoretical framework for understanding the coupled evolution of the physical properties of substellar objects.

\subsection{Comparison of Distance Estimates from Two Methods}
\begin{figure}[htbp]
\centering
\includegraphics[width=\hsize]{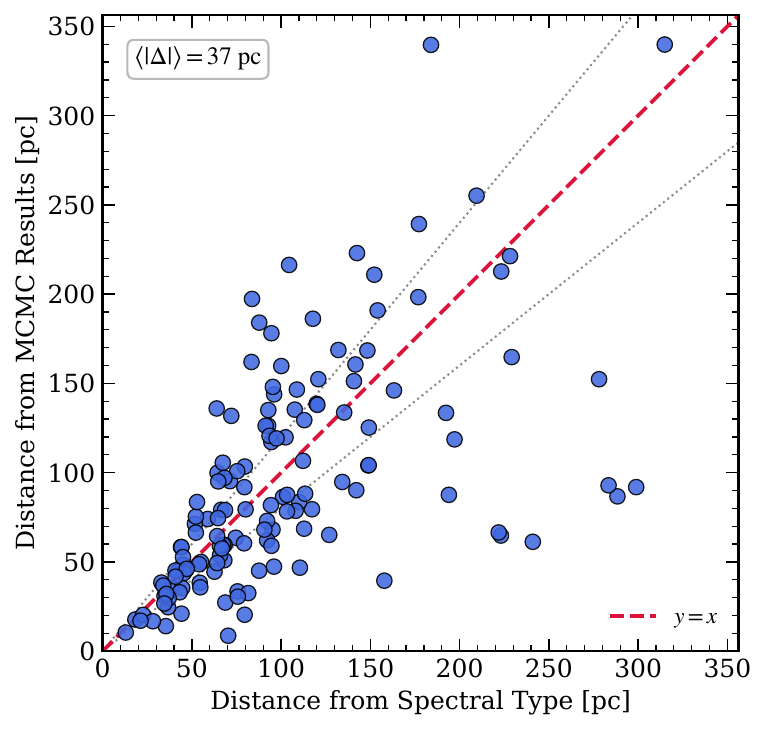}
\caption{Comparison between distances estimated from spectral types and those derived from MCMC fitting. The red dashed line indicates the one-to-one relation ($y=x$), while the grey dotted lines represent deviations of $\pm 20\%$ from this relation. The mean absolute distance difference is $\langle |\Delta d| \rangle = 37$ pc.}
\label{fig:distance}
\end{figure}

We estimated the distances to our T-BDs using two independent approaches: the spectral-type-based method and the MCMC-based method. The former provides an empirical distance estimate based on the relationship between spectral type and absolute magnitude, while the latter derives the distance through fitting the observed data with the adopted theoretical evolutionary and atmospheric models.

For the spectral-type-based method, we estimated the absolute magnitude of each object using the adopted empirical spectral type--absolute magnitude relation \citep{2024RNAAS...8..137S}. By combining the estimated absolute magnitude with the corresponding apparent magnitude, the distance was calculated using the distance modulus relation:
\begin{equation}
d = 10^{(m-M+5)/5}\ {\rm pc},
\end{equation}
where $m$ and $M$ represent the apparent and absolute magnitudes, respectively.

For the second method, the distance was inferred by combining the radius derived from the evolutionary models with the flux-scaling parameter, $(R^2/D^2)$. The posterior samples of the radius and scaling parameter were propagated to obtain the posterior distribution of the distance for each source. We adopted the median of this distribution as the representative distance estimate.

Fig.~\ref{fig:distance} compares the distances obtained from these two methods. The spectral-type-based distances are shown along the horizontal axis, while the MCMC-derived distances are plotted along the vertical axis. The red dashed line represents the one-to-one relation. The dotted lines illustrate the dispersion around this relation. The mean absolute difference between the two distance estimates is approximately 37~pc.

Overall, the two methods show a positive correlation, suggesting that they provide broadly consistent distance estimates for a considerable fraction of our sample. Most sources with relatively small distances are distributed around the one-to-one relation, indicating reasonable agreement between the empirical spectral-type-based estimates and the MCMC results. However, the scatter becomes more pronounced at larger distances, where several sources deviate substantially from the one-to-one relation.

In particular, a number of sources lie above the one-to-one line, indicating that the MCMC method yields larger distances than the spectral-type-based method. Conversely, several sources, especially among those with spectral-type-based distances above approximately 150~pc, lie below the one-to-one relation, suggesting systematically smaller distances from the MCMC analysis for these objects. This asymmetric distribution of the deviations indicates that the differences between the two methods are not entirely random and may be related to the assumptions and limitations of the adopted models.

The discrepancies between the two distance estimates may arise from several factors. The spectral-type-based method relies on empirical spectral type--absolute magnitude relations, which are subject to intrinsic scatter and uncertainties in spectral classification. In addition, variations in atmospheric properties, such as surface gravity, metallicity, and cloud properties, can introduce additional dispersion in the absolute magnitudes of brown dwarfs with similar spectral types. On the other hand, the MCMC-derived distances depend on the adopted evolutionary and atmospheric model prescriptions, as well as the quality of the observational data and the degeneracies among the fitted parameters. These differences may become more significant for relatively distant and intrinsically faint sources, for which the observational constraints are generally weaker.

\section{Discussion}
\label{sec:Discussion}

\subsection{Comparison with the Euclid Q1 Brown Dwarf Sample}

Previous studies based on the \textit{Euclid} Q1 data identified a total of 18 T-BDs, whereas only seven of these sources are included in our final sample. This difference is mainly associated with the adopted photometric measurements, colour-selection criteria, and subsequent spectroscopic quality requirements.

One important difference between our analysis and that of \citet{kiwy2026search} lies in the photometric measurements used to construct the colour--colour diagrams. They adopted fluxes obtained through a template-fitting approach, whereas we used aperture photometry measured within apertures of $2\,{\rm FWHM}$. Although both approaches have their respective advantages, small systematic differences in the measured fluxes can arise from the adopted photometric methods. Such differences can propagate into the derived colours and consequently affect the positions of individual sources in the colour--colour diagram. In particular, some objects may shift across the boundaries of our adopted colour-selection region and therefore are not recovered by our selection criteria.

After accounting for the differences in the photometric selection, 11 of the 18 T-BDs reported in Q1 fall within our adopted colour-selection region. Among these 11 sources, seven have spectra of sufficient quality to satisfy our subsequent spectroscopic selection requirements and are therefore retained in our final sample. The remaining four sources are excluded because of insufficient spectral quality, which prevents reliable spectroscopic analysis and atmospheric parameter determination.

\subsection{Candidate Brown Dwarf Multiple Systems}

\begin{figure}[htbp]
\centering
\includegraphics[width=\hsize]{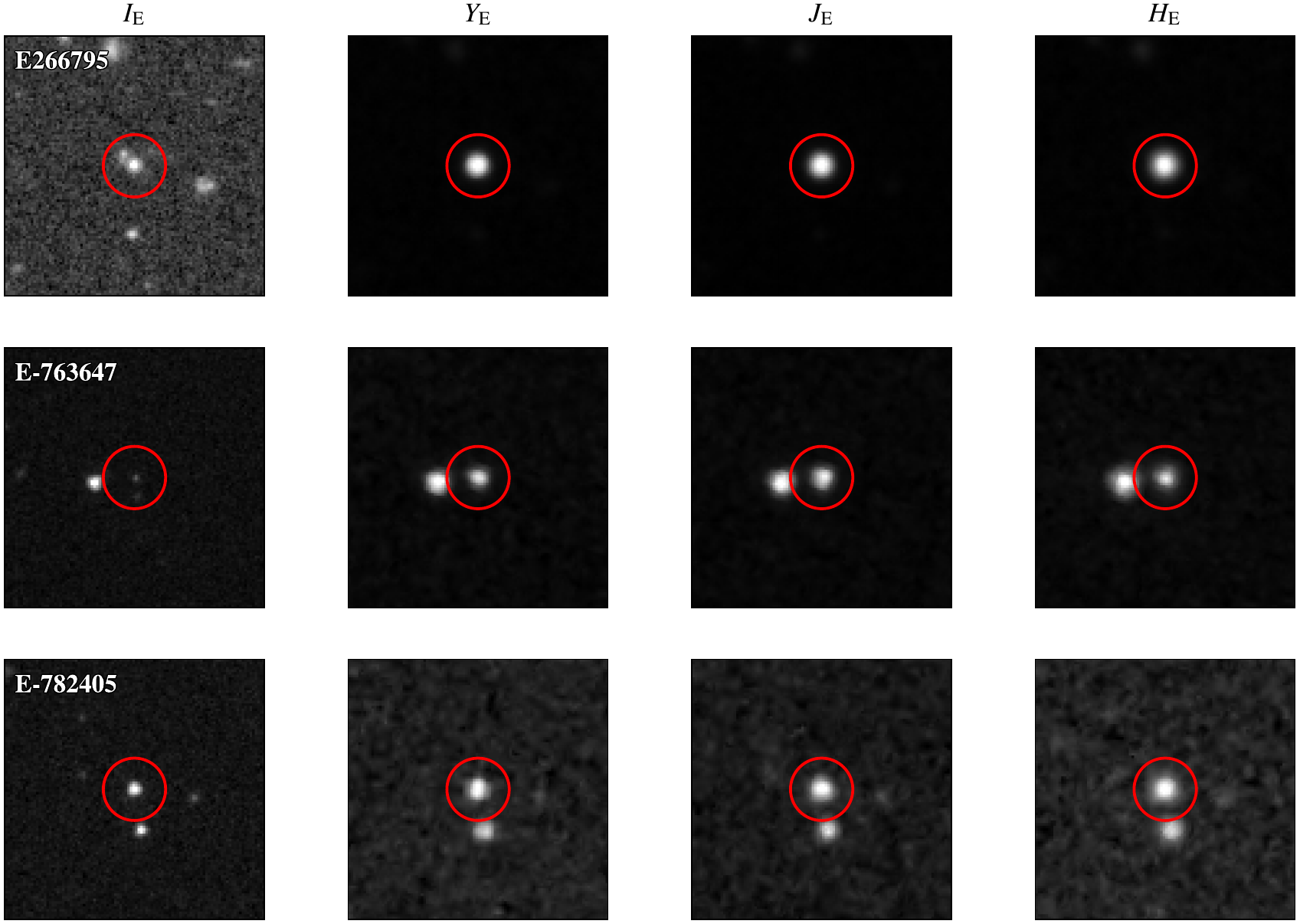}
\caption{Euclid multi-band ($I_{\rm E}$, $Y_{\rm E}$, $J_{\rm E}$, and $H_{\rm E}$) images of three brown dwarf candidates showing nearby sources. The red circles indicate the positions of the primary targets. The multi-band detections of the neighbouring sources in the second and third rows make them promising candidates for potential binary systems, while the neighbouring source in the first system is only clearly detected in the $I_{\rm E}$ band.}
\label{fig:binary_candidates}
\end{figure}

We identified three T-BDs with nearby sources in the Euclid multi-band images, making them potential binary or multiple-system candidates. Fig.~\ref{fig:binary_candidates} presents their cutout images in the $I_{\rm E}$, $Y_{\rm E}$, $J_{\rm E}$, and $H_{\rm E}$ bands. However, the nature of these nearby sources cannot be established from imaging data alone, and further observations are required to determine whether they are physically associated companions or chance-aligned background or foreground objects.

The first source, with E266795 (\texttt{object\_id = 2667957196649082669}), exhibits a relatively distinct nearby source in the $I_{\rm E}$ image. Interestingly, no corresponding source is clearly detected at the same position in the near-infrared bands. This suggests that the neighbouring source may have a relatively blue spectral energy distribution, with its emission being more prominent at optical wavelengths than in the near-infrared. It could therefore be an unrelated foreground or background source rather than a physical companion to the brown dwarf. Other possibilities, including a compact background galaxy or an imaging artefact, cannot be excluded. Consequently, the physical association of this neighbouring source remains uncertain.

The second source E-763647 \texttt{(object\_id = -763647152490907557)} shows a more interesting configuration. Two nearby sources can be distinguished in the $I_{\rm E}$ image, and both appear to have corresponding emission in the near-infrared images. The consistency of the two sources across multiple wavelength bands suggests that they are genuine astronomical sources rather than single-band detection artefacts. E-763647 therefore represents a promising candidate system with one or more companions.

The source E-782405 (\texttt{object\_id = -782405528294682675}) has a nearby source, \texttt{object\_id = -782404715294687006}, at an angular separation of only $1.58^{\prime\prime}$. Both objects are spatially resolved in the \textit{Euclid} multi-band images and exhibit relatively faint VIS emission but substantially stronger near-infrared emission. Independent spectral fitting indicates that both sources show spectral characteristics consistent with T-BDs. Interestingly, both objects also display unusual colours, similar to the blue T-BD population discussed above.

If the colours of these two objects are indeed caused by unresolved white dwarf companions, each source could itself represent an unresolved WD+T-BD system. In that case, if the two spatially resolved objects are also physically associated, the system could represent a hierarchical quadruple configuration consisting of two WD+T-BD binaries in a wider orbit. Such a configuration would be particularly interesting because it would combine two compact-object--brown-dwarf binaries within the same bound system. However, this interpretation remains highly speculative at present.

The estimated distances of the two sources differ by approximately 60~pc, which introduces significant uncertainty regarding their physical association. As shown in Fig.~\ref{fig:distance}, the discrepancies between the spectral-type-based and MCMC-derived distances increase for some sources at larger distances, so the observed distance difference may partly reflect the uncertainties of the adopted distance-estimation methods rather than necessarily indicating that the two objects are unrelated. Given their small angular separation, independently identified T-type spectral characteristics, and similarly unusual colours, this pair is an especially interesting target for follow-up observations.

Resolved brown dwarf binaries have previously been identified through high-resolution imaging, resolved spectroscopy, and astrometric measurements \citep{2009AJ....138.1563B,2012ApJ...745...26B}. Our candidate is particularly interesting because both components are independently classified as T-BDs using the Euclid spectroscopic data. Nevertheless, the relatively large difference between their estimated distances prevents us from establishing a physical association at present. Further follow-up observations are therefore essential to clarify the nature of this pair.

\section{Summary}
\label{sec:Summary}

In this work, we conducted a systematic search for and characterization of T-BDs using the near-infrared spectroscopic and multi-band photometric data from \textit{Euclid} DR1. By combining photometric selection, spectral classification, atmospheric model fitting, and evolutionary analysis, we investigated the atmospheric and fundamental physical properties of the identified objects. Our main findings are summarized as follows.

\begin{enumerate}

\item We constructed a sample of T-BDs using morphological selection criteria, near-infrared colour cuts, and spectroscopic analysis. We identified 141 T-BDs through the characteristic molecular absorption features in their Euclid NISP spectra. Among them, 134 are newly identified candidates, while seven have previously been reported. This demonstrates the capability of Euclid DR1 to expand the known population of T-BDs through its combination of wide-area imaging and near-infrared spectroscopy.

\item We determined the atmospheric parameters of the T-BDs by fitting their spectra with the Sonora Elf Owl and LOWZ model grids, and further estimated their masses, ages, and radii using evolutionary models. The effective temperatures derived from the two atmospheric models show good overall agreement, while other parameters exhibit larger model-dependent differences. The inferred mass--radius--age distributions are broadly consistent with the expected evolutionary trends of brown dwarfs.

\item We identified 21 relatively blue T-BDs with \(I_{\rm E}-H_{\rm E}<2.5\), whose spectra retain the characteristic features of T-BDs. Their SEDs and colour distribution, including the overlap with previously reported WD+BD candidates, suggest that some of these sources may be unresolved WD+T-BD systems. In addition, three T-BDs exhibit nearby sources in the \textit{Euclid} images and may be candidate multiple systems. Further observations are required to confirm the nature of these sources.

\item We compared the distances obtained from empirical spectral type--absolute magnitude relations with those derived from MCMC fitting. The two methods show an overall positive correlation, although noticeable scatter is present, particularly at larger distances. The mean absolute difference between the two estimates is approximately 37 pc, indicating that the inferred distances remain sensitive to the adopted methodology and its underlying assumptions. Independent parallax measurements will be important for further evaluating the reliability of these estimates.

\end{enumerate}

Overall, this work demonstrates the scientific potential of \textit{Euclid} DR1 for the identification and characterization of T-BDs. The combination of near-infrared spectroscopy, atmospheric model fitting, and evolutionary analysis provides a framework for investigating their atmospheric properties and physical evolution. Future spectroscopic and astrometric observations will be essential for improving the accuracy of the derived parameters and confirming the nature of the potential binary systems identified in this study.

\begin{acknowledgements}
      Part of this work was supported by the German
      \emph{Deut\-sche For\-schungs\-ge\-mein\-schaft, DFG\/} project
      number Ts~17/2--1.
\end{acknowledgements}

\bibliographystyle{aa}      
\bibliography{aa_reference} 

\begin{landscape}
\begin{table}[p]
\caption{Multi-band flux densities of the Euclid DR1 T-BDs with $I_{\rm E}-H_{\rm E}<2.5$. The flux densities are estimated by applying template-fit colour corrections to the VIS detection-band total fluxes and are given in $\mu$Jy. The final column lists the colours, $I_{\rm E}-H_{\rm E}$.}
\label{tab:blue_source_flux}
\centering
\scriptsize
\setlength{\tabcolsep}{3pt}
\begin{tabular*}{\linewidth}{@{\extracolsep{\fill}}lrrrrrrrrrrrrrr}
\hline
\noalign{\smallskip}
ID & MegaCam u & HSC g & DECam g & MegaCam r & DECam r & $I_{\rm E}$ & PS1 i & DECam i & HSC z & DECam z & $Y_{\rm E}$ & $J_{\rm E}$ & $H_{\rm E}$ & $I_{\rm E}-H_{\rm E}$ \\
\noalign{\smallskip}
\hline
\noalign{\smallskip}
-395079009422377496 & -- & -- & 2.586 & -- & 4.511 & 4.584 & -- & 5.097 & -- & 5.762 & 5.500 & 5.805 & 4.897 & 0.117 \\
-906552219447679740 & -- & -- & 4.560 & -- & 7.983 & 8.384 & -- & 9.370 & -- & 10.088 & 10.011 & 10.935 & 9.635 & 0.164 \\
-590091462514290240 & -- & -- & 0.907 & -- & 1.646 & 1.943 & -- & 2.197 & -- & 2.434 & 2.330 & 2.642 & 2.340 & 0.280 \\
-218575866587181997 & -- & -- & 0.878 & -- & 1.904 & 2.007 & -- & 2.458 & -- & 2.570 & 2.816 & 3.098 & 2.805 & 0.300 \\
-699150546662377589 & -- & -- & 1.494 & -- & 3.281 & 3.613 & -- & 3.880 & -- & 4.662 & 5.356 & 5.983 & 5.124 & 0.346 \\
-455633390634652996 & -- & -- & 0.468 & -- & 1.582 & 1.821 & -- & 2.449 & -- & 3.060 & 3.304 & 3.795 & 3.466 & 0.645 \\
2778165822569271818 & 0.171 & 0.607 & -- & 1.860 & -- & 2.330 & 3.111 & -- & 3.479 & -- & 4.140 & 4.896 & 4.596 & 0.660 \\
\hline
\end{tabular*}
\end{table}
\end{landscape}
\clearpage
\onecolumn
\begingroup
\scriptsize
\renewcommand{\arraystretch}{1.5}
\setlength{\tabcolsep}{2pt}
\setlength{\LTleft}{0pt}
\setlength{\LTright}{0pt}
\begin{longtable}{@{}lcccccccccccccc@{}}
\caption{Spectral types, atmospheric parameters, physical properties, and distances of the 141 T-BDs identified in this work.}\\
\label{tab:EuclidTBD}\\
\hline\hline
\noalign{\smallskip}
Object ID & \multicolumn{2}{c}{Coordinates} & Spectral type & \multicolumn{5}{c}{Atmospheric parameters} & \multicolumn{3}{c}{Physical properties} & \multicolumn{2}{c}{Distances} & References \\
\cline{2-3}\cline{5-9}\cline{10-12}\cline{13-14}
 & RA & Dec & SpT & $T_{\mathrm{eff}}$ & $\log g$ & $[\mathrm{M/H}]$ & C/O & $\log K_{\mathrm{zz}}$ & Mass & $\log_{10}(\mathrm{Age/yr})$ & Radius & $d_{\mathrm{phot}}$ & $d_{\mathrm{MCMC}}$ & \\
\noalign{\smallskip}
\hline
\noalign{\smallskip}
\endfirsthead
\caption{continued.}\\
\hline\hline
\noalign{\smallskip}
Object ID & \multicolumn{2}{c}{Coordinates} & Spectral type & \multicolumn{5}{c}{Atmospheric parameters} & \multicolumn{3}{c}{Physical properties} & \multicolumn{2}{c}{Distances} & References \\
\cline{2-3}\cline{5-9}\cline{10-12}\cline{13-14}
 & RA & Dec & SpT & $T_{\mathrm{eff}}$ & $\log g$ & $[\mathrm{M/H}]$ & C/O & $\log K_{\mathrm{zz}}$ & Mass & $\log_{10}(\mathrm{Age/yr})$ & Radius & $d_{\mathrm{phot}}$ & $d_{\mathrm{MCMC}}$ & \\
\noalign{\smallskip}
\hline
\noalign{\smallskip}
\endhead
\hline
\endfoot
\hline\hline
\endlastfoot
-177156401572535827 & $17.715640$ & $-57.253583$ & T6.0 $\pm$ 0.99 & $1102^{+47}_{-33}$ & $3.34^{+0.47}_{-0.23}$ & $0.96^{+0.03}_{-0.08}$ & $1.10^{+0.02}_{-0.02}$ & $8.80^{+0.16}_{-0.34}$ & $3.29^{+2.38}_{-1.13}$ & $6.62^{+0.48}_{-0.39}$ & $1.46^{+0.11}_{-0.12}$ & $228.2^{+6.7}_{-6.7}$ & $221.3^{+25.1}_{-22.9}$ & \ldots \\
2640960888609883970 & $264.096089$ & $60.988397$ & T7.0 $\pm$ 0.55 & $640^{+5}_{-7}$ & $4.60^{+0.17}_{-0.09}$ & $-0.56^{+0.11}_{-0.03}$ & $0.67^{+0.03}_{-0.03}$ & $2.22^{+0.54}_{-0.16}$ & $18.70^{+8.80}_{-6.54}$ & $9.07^{+0.40}_{-0.37}$ & $0.98^{+0.08}_{-0.08}$ & $35.4^{+1.9}_{-1.9}$ & $14.0^{+1.2}_{-1.2}$ & \ldots \\
-298244988529432861 & $29.824499$ & $-52.943286$ & T7.0 $\pm$ 0.59 & $655^{+9}_{-8}$ & $5.26^{+0.12}_{-0.11}$ & $-0.12^{+0.06}_{-0.07}$ & $0.54^{+0.07}_{-0.06}$ & $6.93^{+1.13}_{-2.31}$ & $36.26^{+6.18}_{-7.93}$ & $9.73^{+0.19}_{-0.28}$ & $0.84^{+0.05}_{-0.03}$ & $44.2^{+2.3}_{-2.3}$ & $21.0^{+1.2}_{-1.2}$ & \ldots \\
1475151535664106148 & $147.515154$ & $66.410615$ & T6.0 $\pm$ 0.52 & $697^{+3}_{-50}$ & $5.24^{+0.02}_{-0.04}$ & $-0.43^{+0.03}_{-0.05}$ & $0.45^{+0.02}_{-0.03}$ & $8.63^{+0.29}_{-6.51}$ & $38.02^{+6.72}_{-8.90}$ & $9.70^{+0.21}_{-0.31}$ & $0.84^{+0.06}_{-0.03}$ & $68.7^{+2.4}_{-2.4}$ & $27.2^{+4.2}_{-3.8}$ & \ldots \\
-571056276502790798 & $57.105628$ & $-50.279080$ & T5.5 $\pm$ 0.71 & $702^{+37}_{-16}$ & $5.22^{+0.02}_{-0.76}$ & $-0.54^{+0.04}_{-0.43}$ & $0.27^{+0.07}_{-0.06}$ & $-0.85^{+0.31}_{-0.11}$ & $35.23^{+8.72}_{-11.59}$ & $9.60^{+0.28}_{-0.43}$ & $0.86^{+0.09}_{-0.05}$ & $75.6^{+2.8}_{-2.8}$ & $33.5^{+4.8}_{-4.7}$ & $^{1}$ \\
-635916177498194218 & $63.591618$ & $-49.819422$ & T6.0 $\pm$ 0.86 & $704^{+50}_{-4}$ & $5.24^{+0.01}_{-1.34}$ & $-0.49^{+1.23}_{-0.01}$ & $0.84^{+0.00}_{-0.29}$ & $-0.95^{+0.18}_{-0.04}$ & $31.35^{+11.45}_{-13.93}$ & $9.47^{+0.37}_{-0.62}$ & $0.88^{+0.12}_{-0.07}$ & $110.5^{+2.5}_{-2.5}$ & $46.7^{+12.9}_{-10.2}$ & \ldots \\
\end{longtable}
\par\noindent\footnotesize\textit{Notes.} Coordinates are given in the ICRS frame and are reported to six decimal places. Spectral types are given with their classification uncertainties. $T_{\mathrm{eff}}$ is in K, $\log g$ and $[\mathrm{M/H}]$ are in dex, mass is in $M_{\mathrm{Jup}}$, age is expressed as $\log_{10}(\mathrm{Age/yr})$, radius is in $R_{\mathrm{Jup}}$, and distances are in pc. Atmospheric and physical parameter uncertainties represent the 16th and 84th posterior percentiles around the median. Photometric distance uncertainties represent the weighted scatter. References: (1) \citet{2026A&A...711A..37E}. (2) \citet{2025ApJ...991...84D}. (3) \citet{2024A&A...686A.171Z}. (4) \citet{2026A&A...711A..37E}. (5) \citet{2013ApJS..205....6M}. (6) \citet{kiwy2026search}.
\endgroup
\clearpage
\twocolumn

\end{document}